\documentclass[conference]{IEEEtran}
\IEEEoverridecommandlockouts
\usepackage{cite}
\usepackage{amsmath,amssymb,amsfonts}
\usepackage{algorithmic}
\usepackage{graphicx}
\usepackage{textcomp}
\usepackage{xcolor}
\usepackage{orcidlink}
\usepackage{multirow}
\def\BibTeX{{\rm B\kern-.05em{\sc i\kern-.025em b}\kern-.08em
    T\kern-.1667em\lower.7ex\hbox{E}\kern-.125emX}}
\begin{document}

\title{\textbf{RAMAN}: \textbf{R}esource‑efficient \textbf{A}pproxi\textbf{M}ate Posit Processing for \textbf{A}lgorithm–Hardware Co‑desig\textbf{N}

\thanks{This work was supported by the Special Manpower Development Program for Chip to Start-Up (SMDP-C2S) of the Ministry of Electronics and Information Technology (MeitY), Government of India, under Grant EE-9/2/21--R\&D-E. The authors also thank the Department of Science \& Technology for financial support under the INSPIRE PhD Fellowship Award No.~IF230364.
}
}

 \author{
     \IEEEauthorblockN{Mohd Faisal Khan\IEEEauthorrefmark{1}\orcidlink{0009-0008-2235-8341},
     Mukul Lokhande\IEEEauthorrefmark{1}\orcidlink{0009-0001-8903-5159}, Member, IEEE,\\ Santosh Kumar Vishvakarma\IEEEauthorrefmark{1}\orcidlink{0000-0003-4223-0077}, Senior Member, IEEE.}
     \IEEEauthorblockA{\IEEEauthorrefmark{1}NSDCS Research Group, Department of Electrical Engineering, Indian Institute of Technology Indore, India}
   Email: skvishvakarma@iiti.ac.in \textbf{(Corresponding Author)}
}

\maketitle

\begin{abstract}
Edge-AI applications still face considerable challenges in enhancing computational efficiency in resource-constrained environments. This work presents RAMAN, a resource-efficient and approximate posit(8,2)-based Multiply-Accumulate (MAC) architecture designed to improve hardware efficiency within bandwidth limitations. The proposed REAP (Resource-Efficient Approximate Posit) MAC engine, which is at the core of RAMAN, uses approximation in the posit multiplier to achieve significant area and power reductions with an impact on accuracy. To support diverse AI workloads, this MAC unit is incorporated in a scalable Vector Execution Unit (VEU), which permits hardware reuse and parallelism among deep neural network layers. Furthermore, we propose an algorithm–hardware co-design framework incorporating approximation-aware training to evaluate the impact of hardware-level approximation on application-level performance. Empirical validation on FPGA and ASIC platforms shows that the proposed REAP MAC achieves up to 46\% in LUT savings and 35.66\% area, 31.28\% power reduction, respectively, over the baseline Posit Dot-Product Unit (PDPU) design, while maintaining high accuracy (98.45\%) for handwritten digit recognition. RAMAN demonstrates a promising trade-off between hardware efficiency and learning performance, making it suitable for next-generation edge intelligence.
\end{abstract}


\section{Introduction}

Artificial Intelligence (AI) has been a key factor in transforming human life in diverse interconnected ways, such as reshaping work structure and methodology, learning, decision capabilities, structure and view of the world. With AI, the world has seen a significant rise in human productivity, especially collaborative AI for complementary human strength, compensatory human weaknesses, and continuous, quick adaptation for rapidly changing needs\cite{AI-survey}. This has led to the need for real-time data processing at the tip of the end or even autonomously most of the time with the help of deep neural networks (DNNs). On-device implementation is often constrained by real-time performance, resource constraints, and enhanced data privacy, in contrast to cloud execution.

\begin{figure}[!t]
    \centering
    \includegraphics[width=0.95\columnwidth]{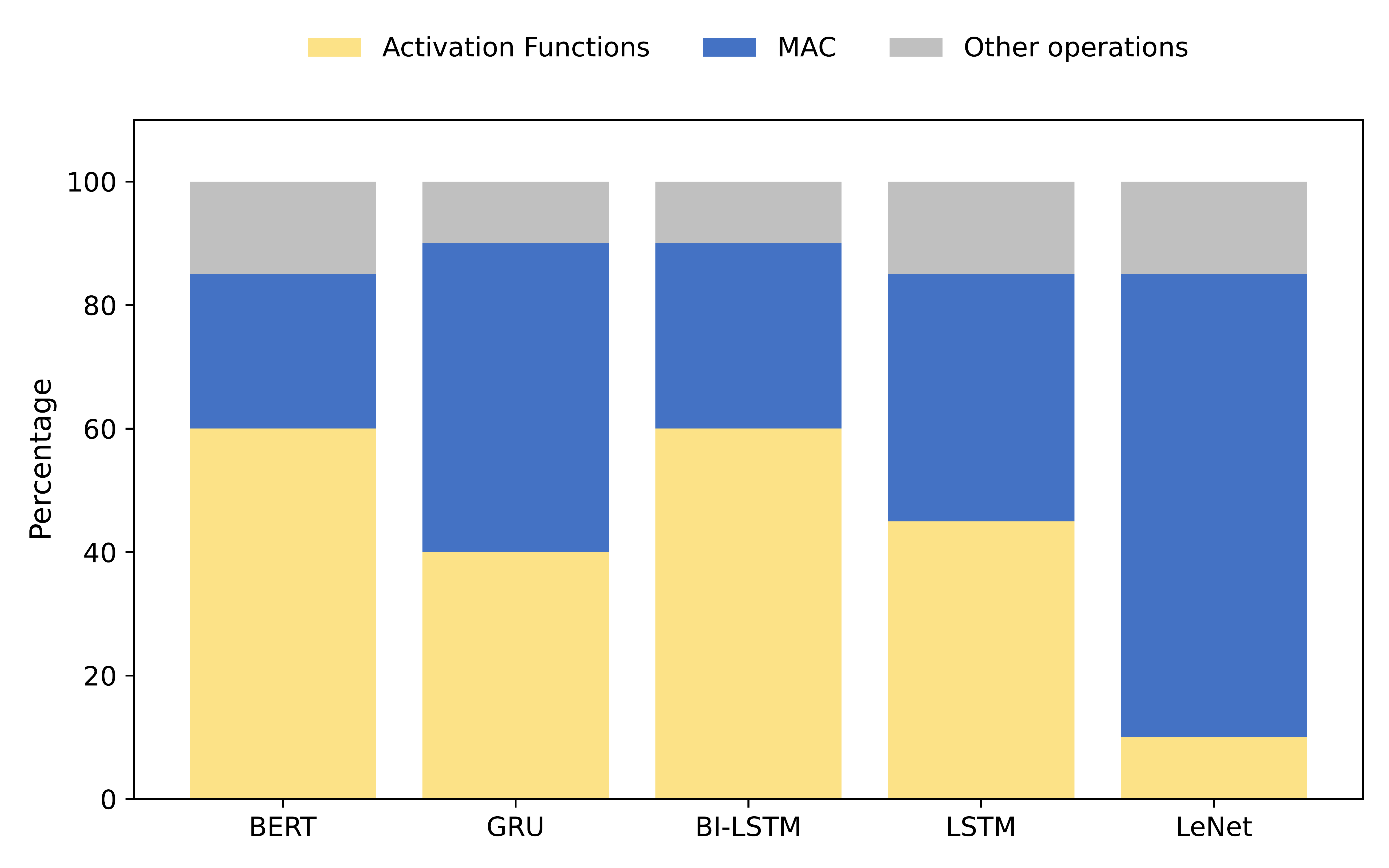}
    \caption{Workload characterization demonstrating implications of Resource-efficient MAC computations for different AI workloads\cite{Flex-PE}.}
    \label{fig:work_char}
\end{figure}

Domain-specific hardware accelerators (DSA) have emerged as one of the solutions for enhanced-performance AI execution\cite{Tutorial}. A complete AI SoC\cite{MAVERIC, Aspen} considers integration of many dedicated accelerators, such as Systolic Array\cite{Flex-PE}, Vector Execution unit\cite{LPRE}, Multi-threaded programmable Linear Algebra Cores, etc\cite{Han-array, Opal}. The in-depth workload characterization (Fig. \ref{fig:work_char}) depicts that the MAC remains dominating computational resources from DNNs to recent models. Another key point to be noticed here is that error-resilient DNNs require more MAC compute operations. As we move to bring more non-linearity, the workloads become relatively more prone to hallucination. The primary objective of this work is to exploit resource efficiency through approximation, memory bandwidth reduction using posit-arithmetic $\mathrm{posit}(n, es)$-where n represents the bit width and es is the exponent size, and apply algorithm-hardware co-design to ensure the application accuracy remains within an acceptable range.  

Prior works \cite {AI-accl1, AI-accl2, Google-TPU} have demonstrated that the accelerator design choices, such as selection of data-width precision, approximation, computational blocks, etc., impact the performance of the complete SoC. For instance, data-width precision\cite{Flex-PE, LPRE} determines the resource efficiency, computational accuracy, and memory bandwidth necessity. The amount of introduced approximation\cite{Quant-MAC, TL, DRALM} provides a trade-off between accuracy and hardware resources (area/power). Prior approximate multipliers\cite{MITCH_TRUNC, HLR-BM, AS_ROBA, RAD-1024, R4ABM, LOBO, AS_ROBA, TL, ALM_SOA, DRALM} such as logarithmic (Mitchell-based and improved variants), Booth/hybrid encoding, RoBA, DRUM, and operand trimming found to provide hardware gains in energy efficiency, area reduction, and computational speed from multipliers\cite{kamal2, kamal3, kamal4, kamal7}. However, most of the above-mentioned approaches tend to have several limitations such as accuracy degradation, non-trivial applications limited to image processing, etc based on approximate adder\cite{kamal0, kamal1, kamal5, kamal6}. This was followed by error compensation mechanisms, which often trade off between accurate and approximate approaches; however, it was not generalized across multiple datasets, bit-precision, or operand distributions. Prior works also limited solutions to specific 8/16-bit or FPGA implementations only, which often fail to address the needs of real-time deployments\cite{Retro,w1,w2,w3}. This approach focuses on algorithm-hardware co-design integration to achieve cumulative gains, such as memory bandwidth, latency, etc. It is based on error metrics (MRED, NMED, WCE) and datasets used for evaluation accuracy with mathematical formulation from \cite{HOAA}. All the factors mentioned above are considered in this work for efficient RAMAN accelerator with Algorithm–Hardware Co-design. 

\begin{figure}
    \centering
    \includegraphics[width=0.875\columnwidth]{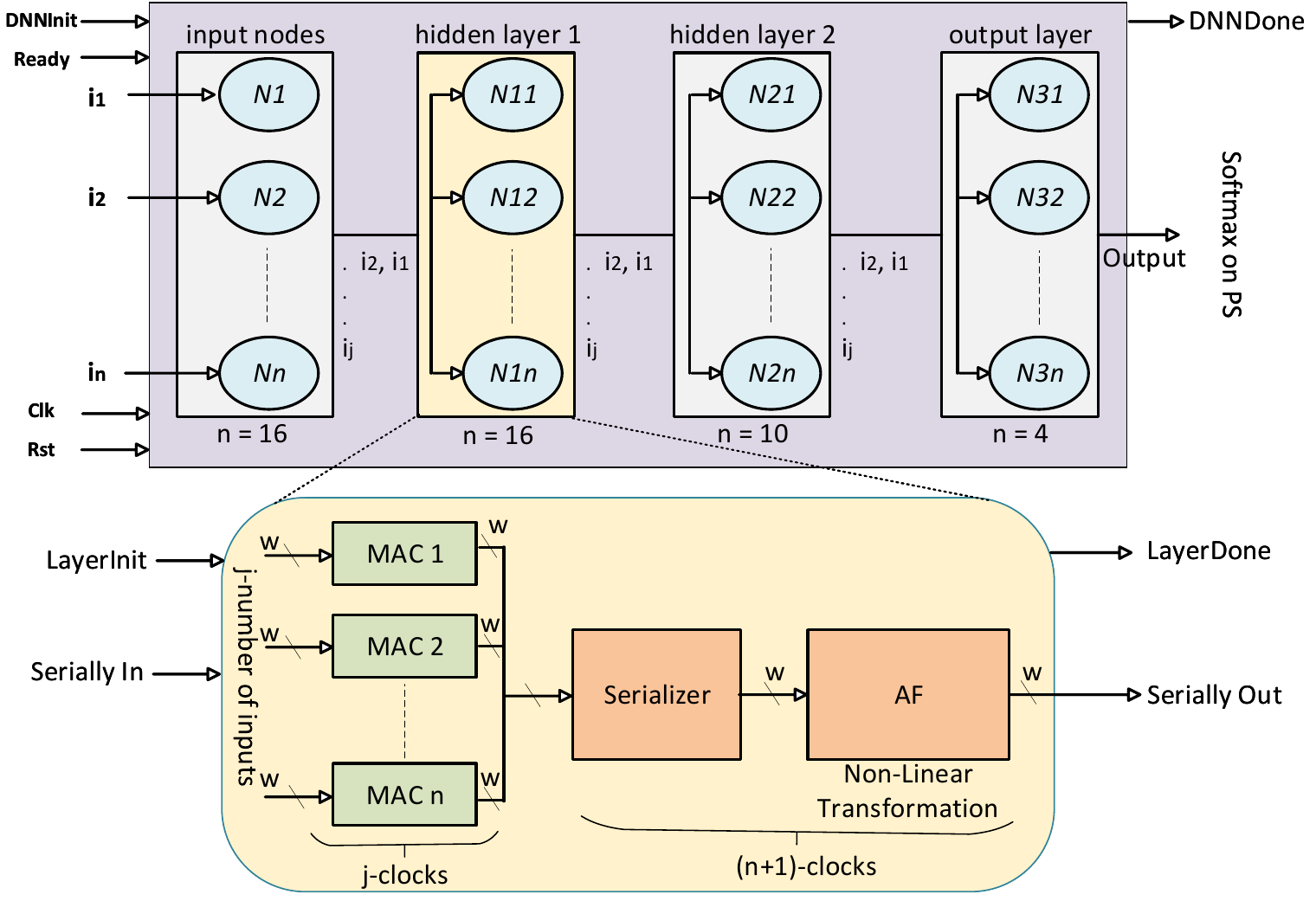}
    \caption{Data multiplexed and hardware reused AI accelerator (HYDRA) architecture with primary emphasis on MAC units, adapted from \cite{GR-Neuro}.}
    \label{fig:layer-multiplexed}
\end{figure}

This work aims to advance the SoTA approaches, with the RAMAN approach for high-performance edge implementation. The key contributions of this work are:

\begin{enumerate}

\item We introduce a Resource-Efficient Approximate $\mathrm{posit}(8,2)$ MAC engine (REAP) for edge AI implementation, with an improvement of up to 46\% in LUT savings, 35.66\% area, and 31.28\% power reduction compared to baseline MAC design.

\item We present a Vector Execution Unit (VEU) integrating the proposed MAC engine as a fundamental compute block for diverse AI accelerators. The proposed vector engine reduces on-chip resource usage and memory bandwidth requirements while improving energy efficiency, enabling more compact and power-efficient AI deployments.

\item We present an algorithm-hardware co-design analysis to evaluate the impact of the proposed approach on application-level performance (accuracy) across diverse AI workloads. 

\end{enumerate}

\begin{figure*}[!t]
    \centering
    \includegraphics[width=0.875\textwidth]{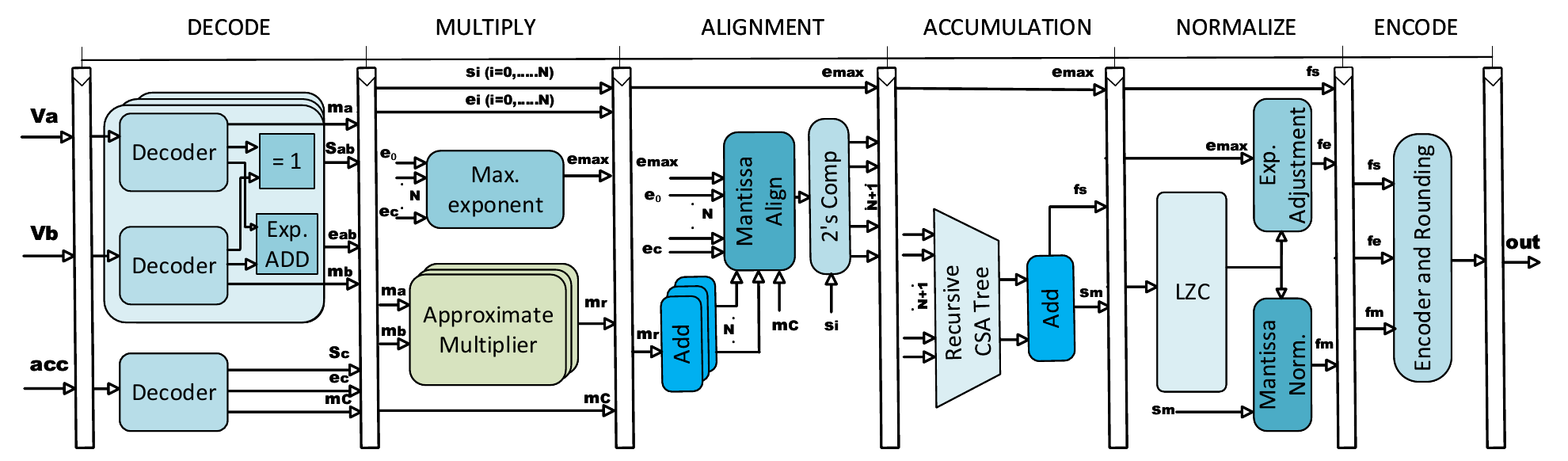}
    \caption{Detailed data-path for the proposed resource-efficient approximate Posit MAC unit.}
    \label{fig:approx-PDPU}
\end{figure*}

\begin{figure}[!b]
    \centering
    \includegraphics[width=0.825\columnwidth]{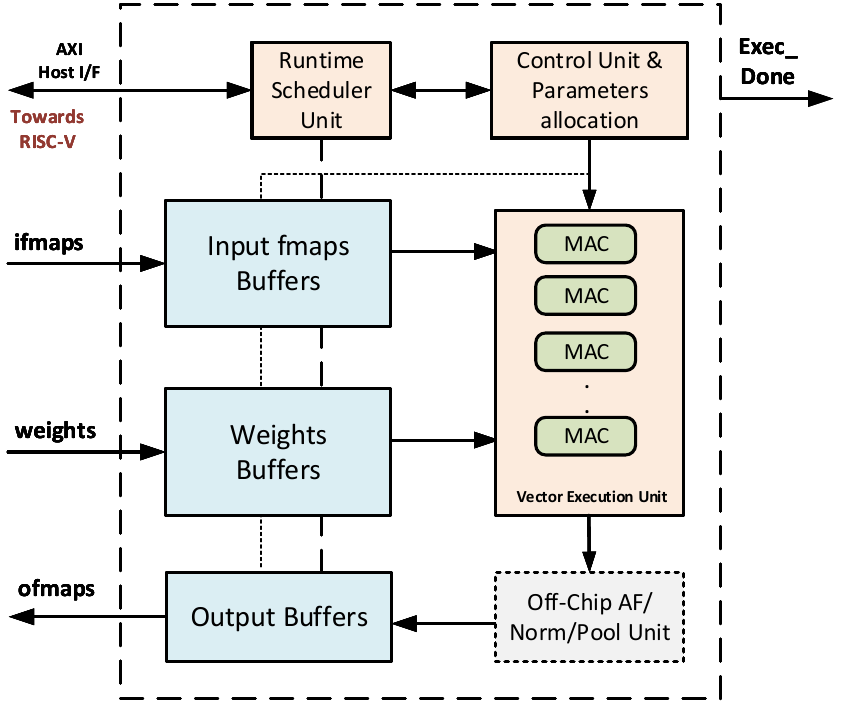}
    \caption{Detailed AI accelerator architecture, showing Vector Execution Unit (VEU)}
    \label{fig:VEU}
\end{figure}

\section{Proposed Work}
Conventional AI accelerators include Matrix Multiplication - systolic array or reused-hardware HYDRA unit (Fig. \ref{fig:layer-multiplexed}). The key focus remains on MAC computations contributing up to 90\% in DNNs and 15.5 GMACs to process a 224x224x3 image with the VGG-16 model.

\subsection{Resource-Efficient Approximate $\mathrm{posit}(8,2)$ MAC}
Posit-8 and Posit-16 have demonstrated superior precision and dynamic range over conventional FP16/BF16 and FP32, respectively. Similarly, DNN accuracy was also observed saving uniform bandwidth by a factor of 2 at least. Fig. \ref{fig:approx-PDPU} depicts the detailed data-path for the proposed resource-efficient approximate $\mathrm{posit}(8,2)$ MAC unit, with a six-stage pipelined architecture. In a low-precision format, it approximates the dot product of two input vectors, Va and Vb, where an approximate multiplier is used to calculate each element-wise multiplication.  As illustrated below, adding the results to the previous output generates a high-precision value out.

\begin{equation}
\text{out} = \text{acc} + \mathbf{V}_a \,\widetilde{\times}\, \mathbf{V}_b
           = \text{acc} + \sum_{i=0}^{N-1} a_i \,\widetilde{\times}\, b_i
\end{equation}

Here, \(\widetilde{\times}\) denotes approximate element-wise multiplication.

The cyclic accumulation occurs with every cycle, after the first five stages are required for the initial pipeline only once. The decode stage extracts the valid sign $s_{ab}$ and exponent $e_{ab}$, and subsequently, the XOR of the output sign is calculated. This work's critical path is multiplying stages, which was conventionally performed with a modified radix-4 Booth multiplier\cite{PDPU} and dominates the overall power and area resources. Thus, we focused on introducing approximation in this multiplier to assess the impact on hardware resources and, subsequently, on accuracy for error-resilient DNN applications. At the same stage, exponent comparison circuitry is utilized for maximum exponent $e_{max}$ calculation. In the alignment stage, the product results are aligned for accumulation based on the difference between the maximum and the corresponding exponent.
Furthermore, for the subsequent addition or subtraction, these are converted into two's complement format. During the accumulation stage, the aligned mantissas are summed with the CSA adder tree to obtain the partial result and the output sign. The output is normalized and exponent re-adjustment based on LZC is done to calculate the final reg/es $f_{e}$ and mantissa $f_{m}$. Furthermore, the rounding and output-packing accumulatively in a single operand are completed during the encode stage.

\subsection{Vector Execution Unit}

The VEU, comprising N-MAC units, is essential in AI accelerators, mainly responsible for accelerating a variety of workloads, including Artificial Neural Networks (ANNs), Multi-Layer Perceptrons (MLPs) within Transformers architectures, Reinforcement Learning models, Generative AI Applications (including those with encoder-decoder structures). Additionally, the same configuration is used to execute Convolutional neural networks (CNNs) and multi-headed attention in Transformers with the help of minor architectural adjustments, such as using im2col transformations in the control unit and configurable parameter allocation and data-feeding mechanisms. 

The detailed accelerator architecture is shown in Fig. \ref{fig:VEU}, emphasizing MAC-heavy VEU. The control unit and Scheduler handle the workload execution in the custom accelerator per the workload parameters, such as model layers, image size, layer type, kernel size, stride, type of pooling, and Activation function. Based on these parameters, the proposed accelerator performs computations in tandem. For instance, convolution layer 1 (C1) of LeNet-5 consists of 6 kernels (5x5) for a 28x28 image, which results in 576 MAC operations per kernel. Given the available shared MAC resources in the VEU, compute cycles would be calculated as 576/N for each kernel, with N-kernel parallel executed in 30 cycles (5 cycles for initial pipeline and 25 cycles for 5x5 kernel MAC compute). The process would repeat 6x 576/N clock cycles for MAC compute of all six kernels in C1, followed by compute cycles necessary in off-chip AF and Max pooling. As the architecture consists of 32x8-b regs, independent of each MAC unit for input, weight and bias, the clock cycles would also be considered for a ping-pong data feed mechanism. This results in 3 clock cycles per MAC unit, each for input, weight, and bias with an AXI-256 interface. Subsequently, 3*N*256 clock cycles feed data for executing VEU once. However, the computational cost associated with the data-feeding mechanism is out of the scope of this work.

\subsection{Algorithm-Hardware co-design}
\begin{figure}[!t]
    \centering
    \includegraphics[width=0.975\linewidth]{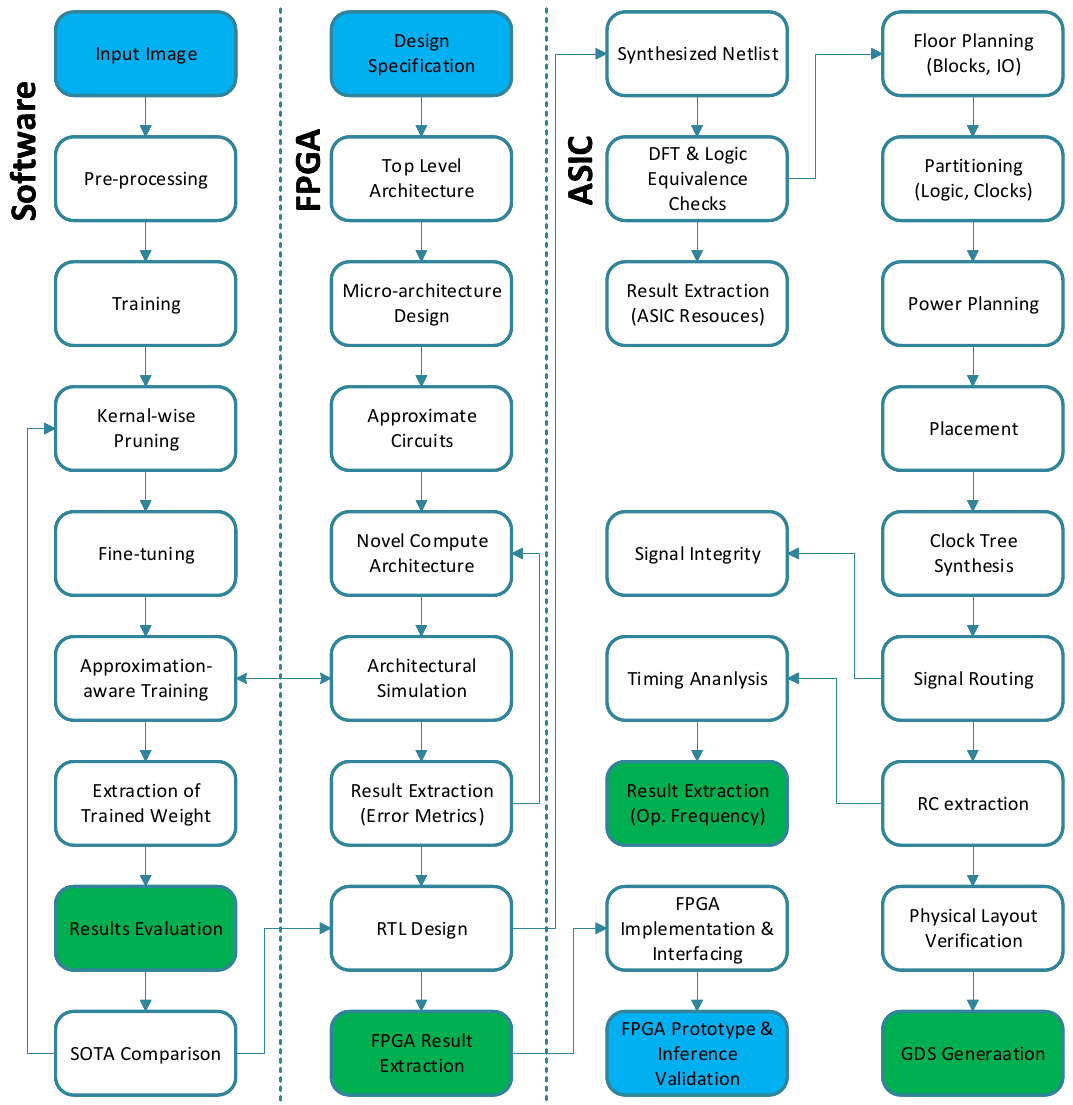}
    \caption{Detailed methodology/workflow describing Algorithm--Hardware co-design utilized in the empirical evaluation of approximate posit processing.}
    \label{fig:codesign}
\end{figure}

The detailed approach to understand algorithm-hardware co-design has been briefed in Fig. \ref{fig:codesign}. The strategy focuses on mutual benefits with posit-quantization on memory-bandwidth reduction, algorithmic performance (accuracy) and respective hardware benefits at the FPGA level and ASIC implementation. The proposed approach during training considers the approximate quantization-aware training for the most-accurate estimation of application accuracy impact derived with specific hardware approximation. Once the estimated accuracy satisfies pre-defined Quality of Results (96.5\%) for edge AI applications, the user can proceed with the respective hardware design and calculate FPGA resources and ASIC parameters for the technology-specific performance evaluation.

The Tiny-YOLOv3 network is used as an example to illustrate approximate quantization-aware training\cite{w4}. Since $\mathrm{posit}(8,2)$ has a greater dynamic range and tapering precision than previous Tiny-YOLOv3 quantization works that use INT8/FP8, it reduces quantization error for small-magnitude weights and activations while maintaining 8-bit storage cost.
This is followed by the mathematical formulation of the suggested $\mathrm{posit}(8,2)$ inference. 
Let $x \in \mathbb{R}^{H \times W \times C}$ be the input image, $\theta = \{W_l, b_l\}_{l=1}^L$ be the set of weights and biases for each of the $L$ layers, $\hat{y} \in \mathbb{R}^K$ be the final output (bounding box coordinates + class probabilities), and $\mathcal{F}$ be the accumulative function corresponding to the convolution, batch-norm, activation, and classification layers in Tiny-YOLOv3.

The quantization function $Q(\cdot)$ is defined as:
\begin{equation}
Q(x) = \mathrm{clip}\!\left( \left\lfloor \frac{x}{\Delta} \right\rceil, q_{\min}, q_{\max} \right) \cdot \Delta
\end{equation}
where the scale factor $\Delta$ is
\begin{equation}
\Delta = \frac{\max(|x|)}{2^{k-1} - 1}
\end{equation}
and
\begin{align}
q_{\min} &= - \left( 2^{k - 1} - 1 \right), \\
q_{\max} &= 2^{k - 1} - 1,
\end{align}
with $k$ denoting the number of quantization bits.

The sensitivity of the $l$-th layer is expressed as:
\begin{equation}
\hat{W}_l = Q(W_l)
\end{equation}

Forward propagation is given by:
\begin{align}
z_l &= \hat{W}_l * \hat{a}_{l-1} + b_l, \\
a_l &= \phi(z_l)
\end{align}
where $*$ denotes convolution, $\phi(\cdot)$ is the activation function, and the previous layer output is quantized as:
\begin{equation}
\hat{a}_{l-1} = Q(a_{l-1}).
\end{equation}

Gradients are computed in FP32 precision. The approximate straight-through estimator (STE) is:
\begin{equation}
\frac{\partial Q(x)}{\partial x} \approx 1 \quad \text{for} \quad x \in [q_{\min} \Delta,\, q_{\max} \Delta].
\end{equation}

Parameter update is then:
\begin{equation}
W_l^{(t+1)} = W_l^{(t)} - \eta \cdot \frac{\partial \mathcal{L}}{\partial \hat{W}_l}
\end{equation}
where $\eta$ is the learning rate and $\mathcal{L}$ represents error in outcomes when comparing the proposed approximate $\mathrm{posit}(8,2)$ and conventional FP32 outputs.

\section{Performance Evaluation}

The empirical assessment for the RAMAN approach involves a comprehensive algorithm-hardware co-design approach, which includes software-based approximate-aware accuracy analysis and hardware architectural design for ASIC performance. The proposed iso-functional behavior was replicated in Python 3.8 with fxp-math library and QKeras 2.6, with the help of Google Colab environment and NVIDIA V100 GPU. Different approximate multipliers and PDPU codes were replicated from the available GitHub repository. We analyzed and integrated the approximate multipliers in the PDPU unit and assessed the NMED error metric as reported in Table \ref{tab:approx_ppa_comp}. Further, functional validation and error metrics were evaluated with a Python framework and Questa-sim simulator, followed by FPGA synthesis and implementation
with AMD Vivado Design Suite. Additionally, all designs were synthesized with Synopsys Design Compiler at CMOS 28nm and corresponding post-synthesis performance metrics reported in Table \ref{tab:approx_ppa_comp}.

\begin{table}[!t]
\caption{Implications of induced approximation on Posit-MAC resources, compared to baseline accurate PDPU.}
\label{tab:approx_ppa_comp}
\renewcommand{\arraystretch}{1.25}
\resizebox{\columnwidth}{!}{%
\begin{tabular}{|l|c|cc|cc|}
\hline
\multicolumn{1}{|c|}{\multirow{2}{*}{\textbf{Design}}} & \multirow{2}{*}{\textbf{Error (\%)}} & \multicolumn{2}{c|}{\textbf{FPGA (VC707)}} & \multicolumn{2}{c|}{\textbf{CMOS 28nm}} \\ \cline{3-6} 
\multicolumn{1}{|c|}{} &  & \multicolumn{1}{c|}{\textbf{LUTs}} & \textbf{Reduction (\%)} & \multicolumn{1}{c|}{\textbf{Area ($\mu$m\textsuperscript{2})}} & \textbf{Power (mW)} \\ \hline
\textbf{PDPU\_Accurate} & 0 & \multicolumn{1}{c|}{979} & 0.00 & \multicolumn{1}{c|}{9579} & 64.83 \\ \hline
\textbf{REAP\_HLR\_BM\cite{HLR-BM}} & 0.01 & \multicolumn{1}{c|}{812} & 17.05 & \multicolumn{1}{c|}{7635} & 50.04 \\ \hline
\textbf{REAP\_AS\_ROBA\cite{AS_ROBA}} & 0.39 & \multicolumn{1}{c|}{736} & 24.82 & \multicolumn{1}{c|}{6999} & 18.24 \\ \hline
\textbf{REAP\_RAD1024\cite{RAD-1024}} & 0.44 & \multicolumn{1}{c|}{793} & 19 & \multicolumn{1}{c|}{6703} & 25.87 \\ \hline
\textbf{REAP\_R4ABM\cite{R4ABM}} & 0.45 & \multicolumn{1}{c|}{634} & 35.24 & \multicolumn{1}{c|}{8471} & 25.32 \\ \hline
\textbf{REAP\_LOBO\cite{LOBO}} & 1.85 & \multicolumn{1}{c|}{798} & 18.5 & \multicolumn{1}{c|}{6639} & 18.48 \\ \hline
\textbf{REAP\_ROBA\cite{AS_ROBA}} & 2.92 & \multicolumn{1}{c|}{644} & 34.2 & \multicolumn{1}{c|}{7323} & 38.49 \\ \hline
\textbf{REAP\_HRALM\cite{TL}} & 7.2 & \multicolumn{1}{c|}{812} & 17.05 & \multicolumn{1}{c|}{6383} & 17.93 \\ \hline
\textbf{REAP\_ALM\_SOA\cite{ALM_SOA}} & 8.06 & \multicolumn{1}{c|}{782} & 20.12 & \multicolumn{1}{c|}{6343} & 20.35 \\ \hline
\textbf{LPRE\_ILM\cite{LPRE}} & 11.84 & \multicolumn{1}{c|}{846} & 13.58 & \multicolumn{1}{c|}{6311} & 17.82 \\ \hline
\textbf{REAP\_DRUM\cite{DRALM}} & 12.43 & \multicolumn{1}{c|}{812} & 17.05 & \multicolumn{1}{c|}{6875} & 43.62 \\ \hline
\textbf{REAP\_MITCH\_TRUNC\cite{MITCH_TRUNC}} & 14.43 & \multicolumn{1}{c|}{795} & 18.8 & \multicolumn{1}{c|}{6307} & 19.24 \\ \hline
\textbf{Proposed} & 6.31& \multicolumn{1}{c|}{526} & 46.28 & \multicolumn{1}{c|}{6163} & 20.28 \\ \hline
\end{tabular}}
\end{table}

\begin{figure}[!t]
    \centering
    \includegraphics[width=0.85\columnwidth]{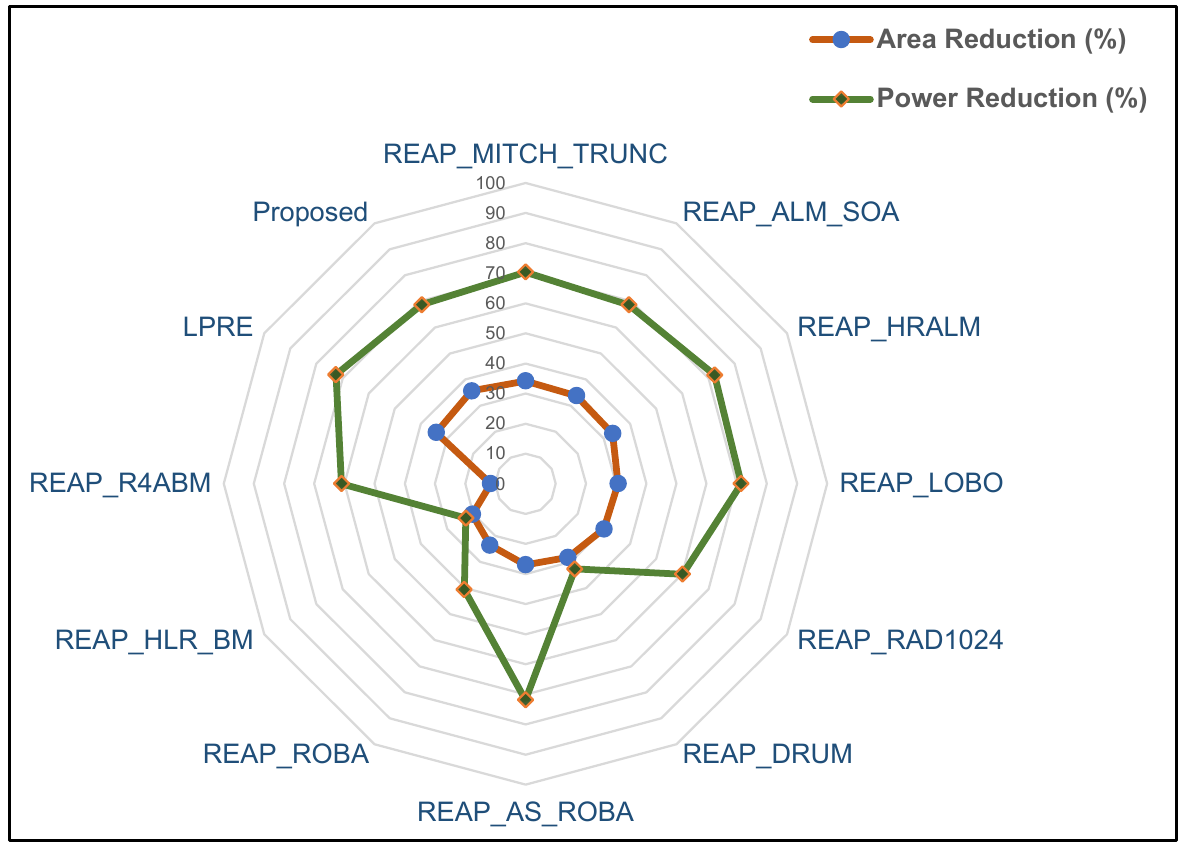}
    \caption{ASIC area and power reduction due to induced approximation, compared to baseline PDPU. Reported at 28nm CMOS tech. node, $V_\mathrm{DD}$ 0.9V, 100 MHz, TT corner, 27\textdegree{}C.}
    \label{fig:ppa_reduction}
\end{figure}

The error metrics, FPGA LUT utilization and ASIC performance for most prior works were re-estimated with iso-experimental factors. We also observed reduced LUT resources in our $\mathrm{posit}(8,2)$ approach (526), compared to BF16 (3670 LUTs) and FP32 (8065 LUTs), which demonstrates a significant reduction. 
The key conclusion to be drawn from this work would be that the integration of DR-ALM\cite{DRALM} into the PDPU unit provides us with an approximate $\mathrm{posit}(8,2)$ multiplier, which has significantly reduced the error of 6.31\% and ASIC resources reduction up to 35.66\% in area and 31.28\% in power compared to the accurate PDPU unit. The proposed work was also found to be superior, in terms of reduction in area and error by 2.35\% and 5.53\% compared to SoTA LPRE\cite{LPRE} unit. The detailed comparison for ASIC area and power-reduction due to induced SoTA approximation approaches is compared in Fig. \ref{fig:ppa_reduction}. This work's key motivation is to inform readers of the possible approximate algorithm-hardware implications trade-off. The relationship was found to be inversely proportional.

\begin{figure}[!t]
    \centering
    \includegraphics[width=0.825\columnwidth]{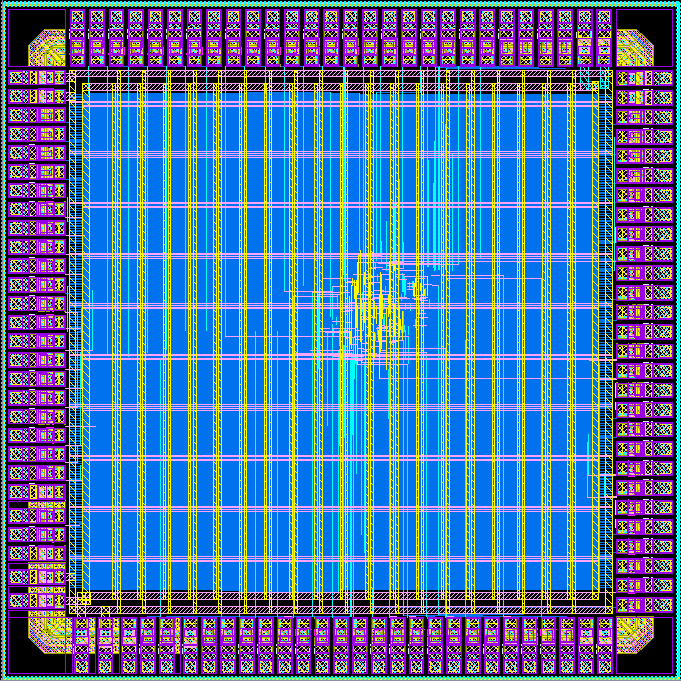}
    \caption{Layout for REAP-MAC engine, including I/O Pad directions with SCL 180nm}
    \label{fig:gds}
\end{figure}

Table \ref{tab:mac_asic} discusses the different SoTA MAC compute units compared in CMOS 28nm. Our design showcases improved area and power compared to other works, which should translate to enhanced energy efficiency and compute density when the proposed MAC compute unit is replaced with modularity. The accelerated performance was also validated with its impact on application accuracy. For simplicity of implementation, we considered the handwritten digit recognition model, which comprises two convolutional layers, each followed by max pooling, two fully connected layers with tanh and a classification softmax layer. The training set and test dataset involve 60k and 10k images from MNIST. The proposed approximate-aware algorithm-hardware co-design framework reports the application accuracy of 98.45\% for this work, compared to 98\% for \cite{MITCH_TRUNC}, 97.2\% for LPRE \cite{LPRE}, 97.12\% for Quant-MAC\cite{Quant-MAC}, 96.47\% for FxP8 DR-ALM\cite{DRALM} and 98.38\% BF16 baseline. We further assessed ASIC implementation of VEU with 256 CUs, individually for LPRE\cite{LPRE} (1.63), PDPU\cite{PDPU} (2.48) and proposed MAC unit (1.57) in  mm\textsuperscript{2} CMOS area. The physical design of proposed REAP-MAC was realized with the indigenous SCL 180 nm process technology and has been submitted for fabrication through a free academic tape-out initiative, as illustrated in Fig. \ref{fig:gds}.

\begin{table}[!t]
\caption{Comparison of ASIC Resources, with SoTA MAC approaches}
\label{tab:mac_asic}
\renewcommand{\arraystretch}{1.15}
\resizebox{\columnwidth}{!}{%
\begin{tabular}{|c|c|c|c|c|c|c|}
\hline
\multirow{2}{*}{\textbf{Design}} & \textbf{Tech.} & \textbf{Voltage} & \textbf{Freq.} & \textbf{Area} & \textbf{Power} & \textbf{PDP} \\ \cline{2-7} 
 & \textbf{nm} & \textbf{V} & \textbf{GHz} & \textbf{mm²} & \textbf{mW} & \textbf{pJ} \\ \hline
\textbf{TCAS-I'25\cite{MPS-FMA}} & 28 & 1 & 0.97 & 0.0276 & 39 & 40 \\ \hline
\textbf{TVLSI'25\cite{Flex-PE}} & 28 & 0.9 & 1.36 & 0.049 & 7.3 & 5.37 \\ \hline
\textbf{VDAT'25\cite{POLARON}} & 28 & 0.9 & 1.86 & 0.011 & 28.2 & 15.2 \\ \hline
\textbf{ISCAS'25\cite{LPRE}} & 28 & 0.9 & 1.12 & 0.024 & 32.68 & 29.2 \\ \hline
\textbf{TCAD'24\cite{DPDAC-TCAD'24}} & 28 & 1 & 1.47 & 0.024 & 82.4 & 56 \\ \hline
\textbf{TCAS-II'22\cite{UVMAC-TCASII'22}} & 28 & 1.05 & 0.67 & 0.052 & 99 & 148 \\ \hline
\textbf{Baseline (PDPU)} & 28 & 1 & 0.63 & 0.009 & 59.3 & 26.7 \\ \hline
\textbf{Proposed} & 28 & 0.9 & 1 & 0.006 & 20.28 & 20.28 \\ \hline
\end{tabular}}
\end{table}

\section{Conclusion \& Future Work}
This work introduces RAMAN, a resource-efficient AI accelerator architecture that leverages approximate $\mathrm{posit}(8,2)$ arithmetic for MAC operations. The VEU architecture enables scalability across diverse AI workloads by facilitating modular deployment and hardware reuse. Through algorithm–hardware co-design, we ensure that the accuracy degradation due to hardware approximation remains minimal, as demonstrated for handwritten digit recognition with less than 0.5\% drop in classification accuracy. Compared to state-of-the-art works, the proposed REAP MAC, by strategically embedding approximation in the multiplier unit achieves up to 46\% LUT savings, demonstrating architectural suitability for real-time edge-AI workloads, with substantial 35.66\% area and 31.28\% power reductions verified through FPGA synthesis and 28 nm ASIC design respectively. Future extensions include evaluating REAP MAC integration in larger AI models and deploying on resource-constrained embedded platforms. We plan to assess the improvement in power consumption due to reduced data movement and incorporate $\mathrm{posit}(8,2)$ precision compared to the BF16 baseline. Following detailed power evaluations, we expect significant improvements in energy savings, particularly for workloads associated with dominant off-chip data movement.

\bibliographystyle{ieeetr}
\bibliography{bib}

\begin{thebibliography}{10}

\bibitem{AI-survey}
X.~Wang, Z.~Tang, J.~Guo, T.~Meng, C.~Wang, T.~Wang, and W.~Jia, ``{Empowering Edge Intelligence: A Comprehensive Survey on On-Device AI Models},'' {\em ACM Comput. Surv.}, vol.~57, Apr. 2025.

\bibitem{Tutorial}
S.~Shao, ``{Tutorial: Domain-Specific Accelerators: From Hardware to Systems},'' in {\em 2024 IEEE International Solid-State Circuits Conference (ISSCC)}, pp.~1--1, 2024.

\bibitem{MAVERIC}
S.~Kim, J.~Zhao, R.~Hsiao, Y.~Chi, V.~Iyer, V.~Jain, B.~Nikolić, and Y.~S. Shao, ``{MAVERIC: A 16nm 72 FPS, 10 mJ/Frame Heterogeneous Robotics SoC with 4 Cores and 13 INT8/FP32 Accelerators},'' in {\em Symposium on VLSI Technology and Circuits}, pp.~1--3, 2025.

\bibitem{Aspen}
K.~Feng, K.~Prabhu, K.~Bartolone, J.~Yu, and P.~Raina, ``{Aspen: A 630 FPS Real-Time Posit-Based Unified Accelerator for Extended Reality Perception Workloads},'' in {\em IEEE Custom Integrated Circuits Conference (CICC)}, pp.~1--3, 2025.

\bibitem{Flex-PE}
M.~Lokhande, G.~Raut, and S.~K. Vishvakarma, ``{Flex-PE: Flexible and SIMD Multiprecision Processing Element for AI Workloads},'' {\em IEEE Transactions on Very Large Scale Integration (VLSI) Systems}, vol.~33, no.~6, pp.~1610--1623, 2025.

\bibitem{LPRE}
O.~Kokane, M.~Lokhande, G.~Raut, A.~Teman, and S.~K. Vishvakarma, ``{LPRE: Logarithmic Posit-enabled Reconfigurable edge-AI Engine},'' in {\em IEEE International Symposium on Circuits and Systems}, pp.~1--5, 2025.

\bibitem{Han-array}
C.~Guo, C.~Wei, J.~Tang, B.~Duan, S.~Han, H.~Li, and Y.~Chen, ``{Transitive Array: An Efficient GEMM Accelerator with Result Reuse},'' in {\em Proceedings of the 52nd Annual International Symposium on Computer Architecture}, ISCA '25, (New York, NY, USA), p.~990–1004, Association for Computing Machinery, 2025.

\bibitem{Opal}
P.-H. Chen, B.~W. Cheng, M.~Oduoza, Z.~Xie, K.~Koul, S.~G. Ravipati, Y.~Mei, R.~Lu, A.~Carsello, M.~Horowitz, and P.~Raina, ``{Opal: A 16nm Coarse-Grained Reconfigurable Array for Full Sparse ML Applications},'' in {\em IEEE Custom Integrated Circuits Conference (CICC)}, pp.~1--3, 2025.

\bibitem{AI-accl1}
Y.~S. Shao, ``{Next-Generation Domain-Specific Accelerators: From Hardware to System},'' in {\em 2024 IEEE Custom Integrated Circuits Conference (CICC)}, pp.~1--5, 2024.

\bibitem{AI-accl2}
M.~Verhelst, L.~Benini, and N.~Verma, ``{How to Keep Pushing ML Accelerator Performance? Know Your Rooflines!},'' {\em IEEE Journal of Solid-State Circuits}, pp.~1--18, 2025.

\bibitem{Google-TPU}
T.~Norrie, N.~Patil, D.~H. Yoon, and et~al., ``{The Design Process for Google's Training Chips: TPUv2 and TPUv3},'' {\em IEEE Micro}, vol.~41, pp.~56--63, Apr. 2022.

\bibitem{Quant-MAC}
N.~Ashar, G.~Raut, V.~Treevedi, S.~K. Vishvakarma, and A.~Kumar, ``{QuantMAC: Enhancing Hardware Performance in DNNs With Quantize Enabled Multiply-Accumulate Unit},'' {\em IEEE Access}, vol.~12, pp.~43600--43614, 2024.

\bibitem{TL}
R.~Pilipović, P.~Bulić, and U.~Lotrič, ``{A Two-Stage Operand Trimming Approximate Logarithmic Multiplier},'' {\em IEEE Transactions on Circuits and Systems I: Regular Papers}, vol.~68, no.~6, pp.~2535--2545, 2021.

\bibitem{DRALM}
P.~Yin, C.~Wang, H.~Waris, W.~Liu, Y.~Han, and F.~Lombardi, ``{Design and Analysis of Energy-Efficient Dynamic Range Approximate Logarithmic Multipliers for Machine Learning},'' {\em IEEE Transactions on Sustainable Computing}, vol.~6, no.~4, pp.~612--625, 2021.

\bibitem{MITCH_TRUNC}
M.~S. Kim, A.~A.~D. Barrio, L.~T. Oliveira, R.~Hermida, and N.~Bagherzadeh, ``{Efficient Mitchell’s Approximate Log Multipliers for Convolutional Neural Networks},'' {\em IEEE Transactions on Computers}, vol.~68, no.~5, pp.~660--675, 2019.

\bibitem{HLR-BM}
H.~Waris, C.~Wang, and W.~Liu, ``{Hybrid Low Radix Encoding-Based Approximate Booth Multipliers},'' {\em IEEE Transactions on Circuits and Systems II: Express Briefs}, vol.~67, no.~12, pp.~3367--3371, 2020.

\bibitem{AS_ROBA}
R.~Zendegani, M.~Kamal, M.~Bahadori, A.~Afzali-Kusha, and M.~Pedram, ``{RoBA Multiplier: A Rounding-Based Approximate Multiplier for High-Speed yet Energy-Efficient Digital Signal Processing},'' {\em IEEE Transactions on Very Large Scale Integration (VLSI) Systems}, vol.~25, no.~2, pp.~393--401, 2017.

\bibitem{RAD-1024}
V.~Leon, G.~Zervakis, D.~Soudris, and K.~Pekmestzi, ``{Approximate Hybrid High Radix Encoding for Energy-Efficient Inexact Multipliers},'' {\em IEEE Transactions on Very Large Scale Integration (VLSI) Systems}, vol.~26, no.~3, pp.~421--430, 2018.

\bibitem{R4ABM}
W.~Liu, L.~Qian, C.~Wang, H.~Jiang, J.~Han, and F.~Lombardi, ``{Design of Approximate Radix-4 Booth Multipliers for Error-Tolerant Computing},'' {\em IEEE Transactions on Computers}, vol.~66, no.~8, pp.~1435--1441, 2017.

\bibitem{LOBO}
R.~Pilipović and P.~Bulić, ``{On the Design of Logarithmic Multiplier Using Radix-4 Booth Encoding},'' {\em IEEE Access}, vol.~8, pp.~64578--64590, 2020.

\bibitem{ALM_SOA}
W.~Liu, J.~Xu, D.~Wang, C.~Wang, P.~Montuschi, and F.~Lombardi, ``{Design and Evaluation of Approximate Logarithmic Multipliers for Low Power Error-Tolerant Applications},'' {\em IEEE Transactions on Circuits and Systems I: Regular Papers}, vol.~65, no.~9, pp.~2856--2868, 2018.

\bibitem{kamal2}
O.~Akbari, M.~Kamal, A.~Afzali-Kusha, and M.~Pedram, ``Dual-quality 4:2 compressors for utilizing in dynamic accuracy configurable multipliers,'' {\em IEEE Transactions on Very Large Scale Integration (VLSI) Systems}, vol.~25, no.~4, pp.~1352--1361, 2017.

\bibitem{kamal3}
S.~Vahdat, M.~Kamal, A.~Afzali-Kusha, and M.~Pedram, ``Tosam: An energy-efficient truncation- and rounding-based scalable approximate multiplier,'' {\em IEEE Transactions on Very Large Scale Integration (VLSI) Systems}, vol.~27, no.~5, pp.~1161--1173, 2019.

\bibitem{kamal4}
H.~Afzali-Kusha, M.~Vaeztourshizi, M.~Kamal, and M.~Pedram, ``Design exploration of energy-efficient accuracy-configurable dadda multipliers with improved lifetime based on voltage overscaling,'' {\em IEEE Transactions on Very Large Scale Integration (VLSI) Systems}, vol.~28, no.~5, pp.~1207--1220, 2020.

\bibitem{kamal7}
M.~Zolfagharinejad, M.~Kamal, A.~Afzali-Khusha, and M.~Pedram, ``Posit process element for using in energy-efficient dnn accelerators,'' {\em IEEE Transactions on Very Large Scale Integration (VLSI) Systems}, vol.~30, no.~6, pp.~844--848, 2022.

\bibitem{kamal0}
M.~Pashaeifar, M.~Kamal, A.~Afzali-Kusha, and M.~Pedram, ``Approximate reverse carry propagate adder for energy-efficient dsp applications,'' {\em IEEE Transactions on Very Large Scale Integration (VLSI) Systems}, vol.~26, no.~11, pp.~2530--2541, 2018.

\bibitem{kamal1}
O.~Akbari, M.~Kamal, A.~Afzali-Kusha, and M.~Pedram, ``Rap-cla: A reconfigurable approximate carry look-ahead adder,'' {\em IEEE Transactions on Circuits and Systems II: Express Briefs}, vol.~65, no.~8, pp.~1089--1093, 2018.

\bibitem{kamal5}
F.~Ebrahimi-Azandaryani, O.~Akbari, M.~Kamal, A.~Afzali-Kusha, and M.~Pedram, ``Block-based carry speculative approximate adder for energy-efficient applications,'' {\em IEEE Transactions on Circuits and Systems II: Express Briefs}, vol.~67, no.~1, pp.~137--141, 2020.

\bibitem{kamal6}
M.~Bahadori, M.~Kamal, A.~Afzali-Kusha, and M.~Pedram, ``High-speed and energy-efficient carry skip adder operating under a wide range of supply voltage levels,'' {\em IEEE Transactions on Very Large Scale Integration (VLSI) Systems}, vol.~24, no.~2, pp.~421--433, 2016.

\bibitem{Retro}
O.~Kokane, G.~Raut, S.~Ullah, M.~Lokhande, A.~Teman, A.~Kumar, and S.~K. Vishvakarma, ``{Retrospective: A CORDIC Based Configurable Activation Function for NN Applications},'' in {\em IEEE Computer Society Annual Symposium on VLSI (ISVLSI)}, vol.~1, pp.~1--6, 2025.

\bibitem{w1}
Z.~Guan, Q.~Liu, and G.~Lin, {\em ACLAM: Accuracy-Configurable Logarithmic Approximate Floating-point Multiplier}, p.~218–223.
\newblock New York, NY, USA: Association for Computing Machinery, 2025.

\bibitem{w2}
S.~Kim, C.~J. Norris, J.~I. Oelund, and R.~A. Rutenbar, ``Area-efficient iterative logarithmic approximate multipliers for ieee 754 and posit numbers,'' {\em IEEE Transactions on Very Large Scale Integration (VLSI) Systems}, vol.~32, no.~3, pp.~455--467, 2024.

\bibitem{w3}
G.~Di~Meo, G.~Saggese, A.~G.~M. Strollo, and D.~De~Caro, ``Approximate mac unit using static segmentation,'' {\em IEEE Transactions on Emerging Topics in Computing}, vol.~12, no.~4, pp.~968--979, 2024.

\bibitem{HOAA}
O.~Kokane, P.~Sati, M.~Lokhande, and S.~K. Vishvakarma, ``{HOAA: Hybrid Overestimating Approximate Adder for Enhanced Performance Processing Engine},'' in {\em 28th International Symposium on VLSI Design and Test (VDAT)}, pp.~1--6, 2024.

\bibitem{GR-Neuro}
G.~Raut, A.~Biasizzo, N.~Dhakad, N.~Gupta, G.~Papa, and S.~K. Vishvakarma, ``{Data multiplexed and hardware reused architecture for DNN accelerators},'' {\em Neurocomputing}, vol.~486, pp.~147--159, 2022.

\bibitem{PDPU}
Q.~Li, C.~Fang, and Z.~Wang, ``{PDPU: An Open-Source Posit Dot-Product Unit for Deep Learning Applications},'' in {\em IEEE International Symposium on Circuits and Systems (ISCAS)}, pp.~1--5, 2023.

\bibitem{w4}
T.~Yu, B.~Wu, K.~Chen, C.~Yan, and W.~Liu, ``Toward efficient retraining: A large-scale approximate neural network framework with cross-layer optimization,'' {\em IEEE Transactions on Very Large Scale Integration (VLSI) Systems}, vol.~32, no.~6, pp.~1004--1017, 2024.

\bibitem{MPS-FMA}
H.~Liu, X.~Lu, X.~Yu, K.~Li, K.~Yang, H.~Xia, S.~Li, and T.~Deng, ``{A 3-D Multi-Precision Scalable Systolic FMA Architecture},'' {\em IEEE Transactions on Circuits and Systems I: Regular Papers}, vol.~72, no.~1, pp.~265--276, 2025.

\bibitem{POLARON}
M.~Lokhande, A.~Jain, and S.~K. Vishvakarma, ``{Precision-aware On-device Learning and Adaptive Runtime-cONfigurable AI acceleration},'' {\em IEEE International Symposium on VLSI Design and Test}, Aug. 2025.

\bibitem{DPDAC-TCAD'24}
H.~Tan, L.~Huang, Z.~Zheng, H.~Guo, Q.~Yang, L.~Shen, G.~Chen, L.~Xiao, and N.~Xiao, ``{A Low-Cost Floating-Point Dot-Product-Dual-Accumulate Architecture for HPC-Enabled AI},'' {\em IEEE Transactions on Computer-Aided Design of Integrated Circuits and Systems}, vol.~43, no.~2, pp.~681--693, 2024.

\bibitem{UVMAC-TCASII'22}
L.~Crespo, P.~Tomás, and N.~Roma, ``{Unified Posit/IEEE-754 Vector MAC Unit for Transprecision Computing},'' {\em IEEE Trans. on Circuits and Syst. II}, vol.~69, pp.~2478--2482, May 2022.

\end{thebibliography}

\end{document}